\documentclass[11pt]{article}

\usepackage{template}
\usepackage{soul}
\usepackage[utf8]{inputenc} 
\usepackage[T1]{fontenc}    
\usepackage{booktabs}       
\usepackage{amsfonts}       
\usepackage{nicefrac}       
\usepackage{microtype}      
\usepackage{lipsum}
\usepackage{fancyhdr}       
\usepackage{graphicx}       
\usepackage{makecell}
\usepackage{tablefootnote}
\usepackage{authblk}
\usepackage{float}
\usepackage[section]{placeins}
\usepackage{lineno}
\usepackage{bm}
\usepackage{amsmath}
\usepackage{xspace}
\usepackage{algorithm}
\usepackage{algpseudocode}
\usepackage{geometry}
\usepackage{xr}

\usepackage{url}            
\PassOptionsToPackage{hyphens}{url}\usepackage{hyperref}
\usepackage{xurl}

\title{
ClimTip-GML: A global bias-corrected and downscaled dataset for assessing impacts of climate tipping events 
}

\author[1,2]{Philipp Hess \thanks{Contact: hess@pik-potsdam.de}}
\author[1,2]{Sebastian Bathiany}
\author[3]{Lucas Ferreira Correa}
\author[4]{Laura C. Jackson}
\author[5]{Casey R. Patrizio}
\author[1,2,6]{Niklas Boers}

\affil[1]{\small Munich Climate Center and Earth System Modelling Group, Department of Aerospace and Geodesy, TUM School of Engineering and Design, Technical University of Munich, Germany}
\affil[2]{Potsdam Institute for Climate Impact Research, Potsdam, Germany}
\affil[3]{Department of Geography, Ludwig-Maximilians University of Munich, Munich, Germany}
\affil[4]{Met Office, Exeter, United Kingdom}
\affil[5]{Institute for Marine and Atmospheric Research Utrecht (IMAU), Utrecht University, Utrecht, Netherlands}
\affil[6]{Global Systems Institute and Department of Mathematics, University of Exeter, Exeter, United Kingdom}

\newcommand{\mpi}{MPI-ESM1-2-HR\xspace}
\newcommand{\cesm}{CESM1-CAM5\xspace}
\newcommand{\hadgem}{HadGEM3-GC31-MM\xspace}

\newcommand{\pr}{pr\xspace}
\newcommand{\tas}{tas\xspace}
\newcommand{\tasmin}{tasmin\xspace}
\newcommand{\tasmax}{tasmax\xspace}
\newcommand{\rsds}{rsds\xspace}
\newcommand{\rlds}{rlds\xspace}
\newcommand{\sfcwind}{sfcWind\xspace}
\newcommand{\hurs}{hurs\xspace}

\begin{document}

\maketitle

\begin{abstract}
    Assessing the impacts of future climate scenarios including tipping events of major Earth system components such as the Amazon rainforest (ARF) or the Atlantic meridional overturning circulation (AMOC), requires accurate and high-resolution simulations. \\
    Here, we present ClimTip-GML, the first globally bias-corrected and downscaled climate dataset for impact assessment of large-scale tipping scenarios, comprising eight key variables at 0.25$^\circ$ spatial resolution from three general circulation models (GCMs): \cesm, \hadgem, and \mpi.
    The dataset includes 100-year-long climate simulations with preindustrial and historical conditions, as well as scenarios at a +2°C warming level with and without tipping transitions of the AMOC or ARF. 
    We apply generative machine learning (GML) techniques trained on reanalysis data to bias-correct and downscale the GCMs in a manner that is physically consistent across space, time, and all eight variables.\\
    Comprehensive validation shows substantially reduced biases, improved small-scale spatial variability, multivariate correlations, and consistent long-term climate responses to the external forcing and tipping events. The results hence permit substantially improved impact assessments of tipping transitions of the ARF and AMOC, directly informing mitigation and adaptation policies.
\end{abstract}

\newpage
\section{Background \& Summary}

Continued anthropogenic global warming poses severe socioeconomic risks. In particular, tipping transitions of Earth system components, such as the Atlantic meridional overturning circulation (AMOC) or the Amazon rainforest (ARF), can have potentially devastating climatic, ecological, and socioeconomic consequences. While uncertainties remain too large to make exact predictions \cite{ben-yamiUncertaintiesTooLarge2024}, increasing signs of stability loss support the concern that we may be approaching such tipping transitions \cite{boultonPronouncedLossAmazon2022, boersDestabilizationEarthSystem2025}, making the assessment of their impacts an urgent task.

To assess these impacts, accurate and high-resolution simulations of tipping scenarios are required. Impact models, e.g., for simulating crop yields or floods, are developed and calibrated with respect to high-resolution observations or reanalysis data and hence require bias-corrected climate simulation input data of similar spatial resolution \cite{frielerScenarioSetupForcing2024}.

General circulation models (GCMs) are the most reliable tool for simulating future climate scenarios \cite{Eyring2016OverviewOrganization}. However, computational constraints limit the horizontal spatial resolution of state-of-the-art GCMs, currently to around 50 -- 200 km \cite{schneiderClimateGoalsComputing2017}, for century-scale simulations. Moreover, systematic errors (biases) in the output of GCMs remain persistent \cite{langeTrendpreservingBiasAdjustment2019, Tian2020ThePrecipitation}.

Regional climate models (RCMs) are often used to dynamically downscale GCMs to higher resolution in a physically consistent manner \cite{giorgiThirtyYearsRegional2019}. However, numerical RCM simulations are computationally expensive and therefore limited to regional domains, and systematic errors in the GCM simulation used as boundary conditions can propagate into the RCM simulation \cite{risserBiasCorrectionDynamical2024}.

To improve the spatial resolution of global simulations, statistical bias correction and downscaling methods can be applied efficiently to the GCM output instead.
For example, the global ISIMIP3b dataset \cite{lange_isimip3b_2021} bias-corrects and downscales a set of GCM simulations from the Coupled Model Intercomparison Project Phase 6 (CMIP6) \cite{Eyring2016OverviewOrganization} for different shared socioeconomic pathways (SSPs) to 0.5$^\circ$ horizontal spatial resolution.
The GloBCORD-HD dataset \cite{yakubu_global_2025} applies the ISIMIP3BASD quantile mapping bias adjustment method to dynamically downscaled CMIP5 simulations for different representative concentration pathways (RCPs) from the CORDEX initiative \cite{giorgiAddressingClimateInformation2009}, providing a quasi-global 0.5$^\circ$ dataset.
The NEX-GDDP dataset \cite{thrasherNASAGlobalDaily2022} bias-corrects and downscales a large set of CMIP6 simulations globally for a selection of SSPs to 0.25$^\circ$ spatial resolution.
Univariate post-processing methods that process each variable and grid cell independently, however, can lead to physical inconsistencies \cite{zscheischlerEffectUnivariateBias2019}.

Generative machine learning (GML) approaches for bias correction and downscaling, e.g. based on diffusion or flow matching \cite{hoDenoisingDiffusionProbabilistic2020, songScoreBasedGenerativeModeling2021, lipmanFlowMatchingGenerative2022}, have emerged as an alternative to process-based RCMs and univariate statistical methods \cite{francoisMultivariateBiasCorrections2020, panLearningCorrectClimate2021, hessPhysicallyConstrainedGenerative2022, hessDeepLearningBiasCorrecting2023, quesada-chaconDownscalingCORDEXDeep2023, hessFastScaleadaptiveUncertaintyaware2025, aichConditionalDiffusionModels2026, mardaniResidualCorrectiveDiffusion2025, schmidtGenerativeFrameworkProbabilistic2025, schillingerEnScaleTemporallyconsistentMultivariate2026, wanRegionalClimateRisk2025}.
These methods can be trained to learn the full high-dimensional distribution of a given high-resolution reference dataset, such as RCM simulations, reanalyses or observations, and generate new samples with accurate spatial, multivariate and temporal dependencies, conditioned on the low-resolution GCM simulation.

Here, we present ClimTip-GML, a novel global dataset for studying impacts of climate tipping transitions, which combines statistical bias correction and GML downscaling to process a comprehensive set of general circulation model simulations from the ClimTip project and as part of the TIPMIP initiative \cite{winkelmannTippingPointsModelling2025, jonesTIPMIPEarthSystem2026}. Our approach is meant to set an example for the bias correction and downscaling of TIPMIP-related simulations in order to enable downstream impact and risk assessments.
ClimTip-GML consists of globally downscaled fields from five 100-year long scenarios produced in the ClimTip project with preindustrial, historical and +2°C warming level conditions, with and without AMOC or ARF tipping, for eight impact variables from three different GCMs (\mpi, \cesm, and \hadgem, see Table \ref{tab:dataset_overview} for an overview). 
A recent generative machine learning method \cite{hessFastScaleadaptiveUncertaintyaware2025}, based on consistency models \cite{song_consistency_2023, songImprovedTechniquesTraining2024} is applied for multivariate downscaling of GCM simulations to 0.25$^\circ$ ($\sim$28 km) horizontal spatial resolution globally. 
Validating ClimTip-GML with respect to key metrics shows substantially reduced biases, accurate representation of small-scale variability, improved multivariate correlations and a well preserved large-scale climate response to external forcing and tipping events.

\section{Methods}

\subsection{General Circulation Model Simulations}

We post-process general circulation model simulations generated with \cesm \cite{meehlClimateChangeProjections2013}, \hadgem \cite{williamsMetOfficeGlobal2018, robertsDescriptionResolutionHierarchy2019}, and \mpi \cite{mullerHigherresolutionVersionMax2018} as part of the ClimTip project \cite{woodControlGlobalTemperature2026} and CMIP6 \cite{Eyring2016OverviewOrganization}. In total, they cover five climate scenarios using consistent simulation protocols from the CMIP6 \cite{Eyring2016OverviewOrganization} and TIPMIP \cite{winkelmannTippingPointsModelling2025, jonesTIPMIPEarthSystem2026} intercomparison projects.

The scenarios are: 1. the pre-industrial control simulation from CMIP (piControl), 2. the historical period simulation from CMIP (historical), 3. the TIPMIP simulation with a global mean temperature of +2°C above pre-industrial, branched off from the idealised CMIP scenario 1pctCO2 (stableT-2K), 4. The stableT-2K scenario, but with additional freshwater hosing to suppress the AMOC (stableT-2K-AMOC), 5. The stableT-2K scenario, but with the Amazon rainforest being removed to imitate an Amazon dieback scenario (stableT-2K-ARF).

The stableT-2K-ARF simulation has not been performed with \hadgem and is not part of our dataset. We hence processed 14 simulations in total.
All GCMs were driven by prescribed CO2 concentrations. We used the 'P-method' described in \cite{woodControlGlobalTemperature2026} to construct the CO2 concentration scenarios required to keep global mean temperature at the target level in the +2°C scenarios.
To force an AMOC decline in the stableT-2K-AMOC simulation, we impose freshwater hosing of 0.3 Sv in the subpolar North Atlantic and Arctic throughout the simulation, following the NAHosMIP protocol \cite{jacksonUnderstandingAMOCStability2023}. In the TIPMIP intercomparison project, this simulation has the identifier tipmip-ocn-p1t2-Ca \cite{swingedouwTIPMIPOCEANExperimentalProtocol2026}.

In the stableT-2K-ARF simulation, we imposed an immediate land cover change from forest to grass in a longitude-latitude box with the limits 17S - 13N and 43W - 81W. This region encloses the majority of the Amazonian forest given the tree cover distribution in all three models. Where different grass types are present in a grid cell before the forest removal, the relative area coverage ratio between the pre-existing grass types is kept. Where no grass type was present, forest was replaced by C4 grass. 
Forest removal happens exactly at the start of the simulation (where it is branched off from the 1pctCO2 simulation). Since prescribed CO2 levels are the same as in stableT-2K, we here only capture the biogeophysical effects of the forest dieback scenario. The radiative and physiological effects of CO2 itself can be assessed by comparing the stableT-2K simulation to piControl, while changes in terrestrial carbon pools can be diagnosed from the GCM output.

The two "tipping" scenarios can be understood as idealized storylines. In reality, it is highly uncertain if, how and at which warming levels or rates and time scales such tipping transitions would occur, and they do not consistently occur in GCMs under moderate warming scenarios. 
In order to still mimic this possibility in the models and study the impacts, we hence interfered in the GCMs in the idealized ways described above (freshwater hosing or instantaneous forest removal). 
These forcings are not intended to represent any realistic future forcing or tipping time scales, but rather the effect of internal processes (in response to greenhouse gas emissions or land use change) that may be misrepresented in models \cite{valdesBuiltStability2011}.

The GCM output for the CMIP scenarios is identical to the public CMIP6 output provided via the Earth System Grid Federation (ESGF), with the following exceptions: 1. The piControl simulation with \mpi is a different realization than the public CMIP6 data. 2. The simulations with \cesm have followed the CMIP5 protocol (not CMIP6, since this model is from an older generation), but are not numerically identical to simulations from the CMIP5 archive since they were running on a different machine and version CESM1.2 was used in here. Further, a different piControl simulation was used as initial conditions for CESM. All realizations of the CMIP6 simulations are r1i1p1f1, except the historical simulation with \hadgem, which is realization r1i1p1f3. Monthly averages of the raw GCM output from these simulations (except the historical simulations) as well as the CO2 forcing and a documentation are available on zenodo \cite{bathianyClimTipESMSimulations2026}. 

Eight key climate impact variables at daily temporal resolution are selected. This selection enables the use of our downscaled dataset for impact model simulations, e.g. with standalone land surface models. The variables are (with CMOR variable name convention in brackets): total precipitation (\pr), daily min, max and mean 2-meter air temperature (\tasmin, \tasmax, and \tas), near-surface wind speed (\sfcwind), short and long-wave downwelling radiation (\rsds, \rlds) and near-surface relative humidity (\hurs).
The native horizontal spatial resolution of the GCM output is 0.94$^\circ \times $ 1.25$^\circ$ in latitude and longitude direction respectively for \cesm, 0.56$^\circ \times $ 0.83$^\circ$ for \hadgem, and 0.94$^\circ \times $ 0.94$^\circ$ for \mpi. 
While the +2°C simulations already have a length of 100 years, we here only use the last 100 years of the historical simulation, which means years 1915 -- 2014 for \mpi and \hadgem, and years 1906 -- 2005 for \cesm (following the CMIP5-protocol, see above). We use these historical simulations as a reference for bias correction. 

\subsection{ERA5 Reanalysis}

ERA5 reanalysis \cite{hersbachERA5GlobalReanalysis2020} from 1979 -- 2023 serves as a reference (``ground truth'') for bias correction and downscaling, with a horizontal spatial resolution of 0.25$^\circ \times $0.25$^\circ$.
ERA5 has been downloaded from the Copernicus Climate Data Store at hourly resolution, which we aggregate to match the daily resolution of the GCM output. The variables \tasmin and \tasmax are the daily minimum and maximum of the hourly 2-meter temperature, and sfcWind is the daily mean of the hourly wind speed.
All GCM simulations are downscaled to the same latitude-longitude-grid of the ERA5 reanalysis reference.
For training the consistency model that is used for downscaling (see below), we split the ERA5 data into a training set 1979 -- 2004, a validation set 2005 -- 2010 and a test set 2011 -- 2023.

\subsection{Data Processing Overview}

We apply the following processing steps to the raw GCM simulations.
We correct systematic errors (biases) at the large (coarse) scale of the GCMs with quantile delta mapping (QDM) \cite{cannon_bias_2015}, applying QDM for each grid cell, day of the year, and variable using empirical distributions (see Section \ref{sec:method_bias_correction} for details).
After the QDM-correction, we interpolate the GCM simulations to the high-resolution ERA5 grid using bilinear interpolation. This only increases the number of grid points and does not yet resolve small subgrid-scale spatial patterns.
Following ISIMIP3BASD \cite{langeTrendpreservingBiasAdjustment2019}, we represent the temperature variables as \tas, the daily temperature range (\tasmax $-$ \tasmin) and its skewness ((\tas $-$ \tasmin) $/$ (\tasmax $-$ \tasmin)), which ensures \tasmin $\leq$ \tas $\leq$ \tasmax after bias correction and downscaling. We apply a log-transform to precipitation and the temperature range, a logit-transform to the skewness, and standardize all variables in the GCM simulations and ERA5 training data by subtracting the mean and dividing by the standard deviation at each grid cell. 
Following the SDEdit-based diffusion-bridge downscaling approach \cite{mengSDEditGuidedImage2022, bischoff_unpaired_2024, hessFastScaleadaptiveUncertaintyaware2025}, we first add noise to the interpolated and standardized GCM data to remove unrealistically blurry patterns up to a spatial scale that is consistent with the coarse GCM resolution.
A generative consistency model, trained as a dynamical multivariate emulator of the ERA5 reanalysis, is then applied following \cite{hessFastScaleadaptiveUncertaintyaware2025} to generate small-scale patterns from the noise added to the GCM fields (see section \ref{sec:method_downscaling} for details).
Afterwards, we reverse the standardization of the downscaled GCM output back to physical units and reconstruct tasmin and tasmax.
To ensure non-negativity of the variables total precipitation (pr), short and long-wave radiation (rsds, rlds), and wind speed (sfcWind), we clamp them to zero. During sampling, precipitation and the temperature range are limited to 600 mm/day and 40 K, respectively.
We further apply a binary mask to the short-wave radiation (rsds) computed from ERA5 climatology, setting all values near the poles to zero. 

\subsection{Bias Correction Methodology}
\label{sec:method_bias_correction}
We use the historical GCM simulations to learn each model's bias with respect to the ERA5 reanalysis, following \cite{hessFastScaleadaptiveUncertaintyaware2025, aichConditionalDiffusionModels2026}. We bilinearly interpolate the ERA5 data to the coarse resolution of the GCM as a reference. The empirical cumulative distribution functions used in QDM are fitted over the ERA5 data from 1979 to 2000 and the historical GCM simulations from the same period, for each variable, grid cell and day of the year, using 500 quantiles. Once fitted, the QDM correction is applied to the full simulation for each GCM and scenario. Since QDM only corrects distributions in time for each grid cell and variable in isolation, this step breaks spatio-temporal and cross-variable consistency. We correct for this by applying the downscaling method. 

\subsection{Downscaling Methodology}
\label{sec:method_consistency_model}
\subsubsection{Consistency Models}

The downscaling method follows the consistency model (CM) approach demonstrated in \cite{hessFastScaleadaptiveUncertaintyaware2025}. Consistency models learn a function with a neural network that 
maps to the same training data (ERA5) sample $\mathbf{x}$ for different noise levels, 
\begin{equation*}
  \mathbf{x}_t = \mathbf{x} + \mathbf{\epsilon}_t, \quad \mathbf{\epsilon}_t \sim \mathcal{N}\left(\mathbf{0}, \sigma_t^2 \mathbf{I}\right),  
\end{equation*}
where the consistency function is,
\begin{equation*}
\mathbf{f}: (\mathbf{x}_{t}, \sigma_t) \mapsto \mathbf{x}, 
\end{equation*}
for every $t\in[t_{\text{min}}, t_{\text{max}}]$, where $\textbf{x}_{t_{min}} \sim p_{\text{data}}(\textbf{x})$ and $\textbf{x}_{t_{\text{max}}} \sim \mathcal{N}\left(\mathbf{0}, \sigma_{t_{\text{max}}}^2 \mathbf{I}\right)$.  
A sample $\textbf{x}$ represents a global high-resolution ERA5 field with $N_{\text{lat}}$ and $N_{\text{lon}}$ grid cells at a given time step and $N_{\text{var}}$ variables, i.e. with dimension $\mathbf{x} \in \mathbb{R}^{N_{\text{var}} \times N_{\text{lat}} \times N_{\text{lon}}}$.
We model temporal dependencies autoregressively by adding the two previous time steps as inputs: 
\begin{equation*}
\mathbf{x}^{n+1} = f_\theta(\mathbf{x}_t^{n+1}, \mathbf{x}^{n}, \mathbf{x}^{n-1}, \sigma_t),
\end{equation*}
where $\mathbf{x}^{n}$, $\mathbf{x}^{n-1}$ are fields of the current and previous time steps. 
The daily top-of-atmosphere insolation is provided as an additional input.
The subscript $t$ denotes the diffusion (noise schedule) time and the superscript $n$ the daily time index. 
After training on ERA5, we use the SDEdit guidance method \cite{bischoff_unpaired_2024, hessFastScaleadaptiveUncertaintyaware2025} to downscale the three different GCMs with a single neural network in a scale-adaptive manner. The training and downscaling techniques are described in the following. 

\subsubsection{Training}

We use the diffusion time discretization proposed in \cite{songImprovedTechniquesTraining2024} with noise scales $\sigma_{\mathrm{\min}} = \sigma_1 < \sigma_2 < ... < \sigma_N = \sigma_{\mathrm{max}}$, and a training schedule  
\begin{equation*}
\sigma_i = \left(\sigma_{\text{min}}^{1/\rho} + \frac{i-1}{N_{\text{train}}(k)-1}(\sigma_{\text{max}}^{1/\rho} - \sigma_{\text{min}}^{1/\rho}) \right)^\rho, \quad i\in [1,N_{\text{train}(k)}],
\end{equation*}
where $k$ is the current training step and $N_{\text{train}}(k) = \min(s_0 2^{[k/K']}, s_1) + 1$ with $K' = \frac{K}{\log_2(s_1/s_0)+1}$, where $K$ is the total number of training steps. See Table \ref{tab:cm_training} for an overview of the parameters used.
The previous time steps are perturbed with noise during training \cite{chenDiffusionForcingNexttoken2024} and with a noise standard deviation of 0.3 during inference, which stabilizes the autoregressive rollout.
We use an exponential moving average of the network weights and a learning rate schedule with warm-up and cosine decay.
We sample a diffusion noise scale $\sigma_i$, with $i \sim p(i)$ according to  \cite{songImprovedTechniquesTraining2024}
\begin{equation*}
p(i) \propto \frac{1}{2} \left[ \mathrm{erf} \left( \frac{\log(\sigma_{i+1}) - P_{\mathrm{mean}}}{\sqrt{2}P_{\mathrm{std}}}\right) - \mathrm{erf} \left( \frac{\log(\sigma_{i}) - P_{\mathrm{mean}}}{\sqrt{2}P_{\mathrm{std}}}\right) \right].
\end{equation*}
Similar to recent work on generative modelling via proper scoring rules \cite{aletOperationalTropicalCyclone2026, stock_swift_2025, schillingerEnScaleTemporallyconsistentMultivariate2026}, we combine the consistency loss with a energy score (ES) regularization,
\begin{equation}
\mathcal{L}(\mathbf{\theta}) = \alpha\mathcal{L}_{\text{CM}}(\mathbf{\theta}) + (1-\alpha) \mathcal{L}_{\text{ES}}(\mathbf{\theta}),
\end{equation}
where $\alpha$ is a hyperparameter. 
The consistency loss term is \cite{songImprovedTechniquesTraining2024},
\begin{equation}
    \mathcal{L}_{\text{CM}}(\theta) =  \mathrm{E}_{i,n}\left[\lambda(\sigma_i) d \left(\mathbf{f}_{\mathbf{\theta}}(\mathbf{x}^{n+1}_{t_{i+1}}, \mathbf{x}^{n}, \mathbf{x}^{n-1}, \mathbf{x}^{\text{stat}}, \sigma_{i+1}), \mathbf{f}_{\mathbf{\theta}}(\mathbf{x}^{n+1}_{t_i}, \mathbf{x}^{n}, \mathbf{x}^{n-1}, \mathbf{x}^{\text{stat}}, \sigma_i) \right)\right],
    \label{eq:cm_loss}
\end{equation}
where spatial fields at time $n$ are sampled uniformly, $n \sim \mathrm{U}[2, N_{\text{time}}-1]$, from the ERA5 reanalysis with length $N_{\text{time}}$. 
The pseudo-Huber loss $d(\mathbf{x}, \mathbf{y}) = \sqrt{||\mathbf{x} - \mathbf{y}||^2_2 + c^2} - c$ is used as the distance measure with a weighting $\lambda(\sigma_i) = 1 /(\sigma_{i+1} - \sigma_i)$ following \cite{songImprovedTechniquesTraining2024}.
The second ES loss term is computed using an ensemble of $M=10$ predictions,
\begin{equation}
    \mathcal{L}_{\text{ES}}(\theta) = \mathrm{E}_{n}\mathrm{E}_{\hat{\mathbf{x}}_{1:M}^{n+1} \sim \mathbf{f}_{\theta}(\mathbf{\epsilon}_{\text{max}}, \mathbf{x}^{n}, \mathbf{x}^{n-1}, \mathbf{x}^{\text{stat}}, \sigma_{\text{max}})} \left[\text{ES} \left(\mathbf{x}^{n+1}, \hat{\mathbf{x}}^{n+1}_{1:M}\right)\right],
\end{equation}
where the ES is defined as \cite{gneitingStrictlyProperScoring2007}
\begin{equation}
\text{ES} \left(\mathbf{x}^{n+1}, \hat{\mathbf{x}}^{n+1}_{1:M} \right)  =  \frac{1}{M}\sum_{i=1}^{M} \left\lVert \hat{\mathbf{x}}_i^{n+1} - \mathbf{x}^{n+1} \right\rVert^\beta_2 - \frac{1}{2M(M-1)}\sum_{i=1}^{M}\sum_{j=1}^{M} \left\lVert \hat{\mathbf{x}}_i^{n+1} - \hat{\mathbf{x}}_j^{n+1} \right\rVert^\beta_2,
    \label{eq:es_loss}
\end{equation}
with $\beta=1$.
We parameterize the CM following \cite{songImprovedTechniquesTraining2024, karrasElucidatingDesignSpace2022a}, with 
\begin{equation}
    f_{\bm{\theta}}(\mathbf{x}_t^{n+1}, \mathbf{x}^{n}, \mathbf{x}^{n-1}, \mathbf{x}^{\text{stat}}, \sigma_t) = c_{\mathrm{skip}}(\sigma_t) \mathbf{x}_t^{n+1} + c_{\mathrm{out}}(\sigma_t) F_{\bm{\theta}}(\mathbf{x}_t^{n+1}, \mathbf{x}^{n}, \mathbf{x}^{n-1},  \mathbf{x}^{\text{stat}}, \sigma_t),
    \label{eq:parameterization}
\end{equation}
where $F(\cdot)$ is a UNet with parameters $\bm{\theta}$. The diffusion time information is transformed using a sin-cosine positional embedding in the network. Following \cite{songImprovedTechniquesTraining2024}, the coefficients $c_{\mathrm{skip}}$ and $c_{\mathrm{out}}$ are defined as
\begin{equation}
c_{\mathrm{skip}}(\sigma_t) = \frac{\sigma^2_{\mathrm{data}}}{(\sigma_t-\sigma_{\mathrm{min}})^2 + \sigma^2_{\mathrm{data}}}, \qquad  c_{\mathrm{out}}(\sigma_t) = \frac{\sigma_{\mathrm{data}} ( \sigma_t - \sigma_{\mathrm{min}})}{\sqrt{\sigma^2 + \sigma^2_{\mathrm{data}}}}.
\label{eq:cm_coeff}
\end{equation}
We use patch-diffusion \cite{wangPatchDiffusionFaster} to improve the efficiency of the neural network training. During training we randomly crop the neural network target and inputs in latitude and longitude direction to $2^k$ grid cells with $k\in [7,8,9]$, which allows for a larger and more expressive neural network (around 250M parameters) that would otherwise not fit in the 80GB GPU memory of the employed NVIDIA H100 when processing global fields. It also acts as a data augmentation and regularization.
The approach requires spatial coordinates as additional inputs to the neural network that we provide using an embedding \cite{watt-meyer_ace_2023} and are denoted above in Eq.~\ref{eq:cm_loss} as $\mathbf{x}^{\text{stat}}$ in the CM input.
We use fixed boundary conditions in the UNet neural network \cite{songScoreBasedGenerativeModeling2021} during training. During inference with global fields we pad the data with 32 grid cells across the date line boundary and crop the final result to the target grid dimensions to avoid boundary effects at the date line.
A summary of all the hyperparameters is given in Table \ref{tab:cm_training}.

\begin{table}[H]
 \caption{Consistency model UNet architecture and training parameters.}
  \centering
  \begin{tabular}{ll|ll|ll}
    \toprule
    \textbf{Architecture}   &        &   \textbf{Noise}          &          & \textbf{Training}       &               \\
    \midrule
    Input channels          & 28     &   $\sigma_{\text{min}}$   & $0.0002$ & Batch size              & 2             \\
    Output channels         & 8      &  $\sigma_{\text{max}}$    & $80$     & Learning rate    & $10^{-4}$     \\
    Resnet blocks           & 2      &  $\sigma_{\text{data}}$   & $0.5$    & LR warm-up             & 5-epoch  \\
    Attention blocks        & 1      &  $\rho$                   & 7        & Optimizer               & AdamW         \\
    Channel multiplier      & (1,2,4,8) &$P_{\text{mean}}$       & -1.1     & Epochs                  & 250           \\
    Channel number          & 128    & $P_{\text{std}}$          & 2.0      & EMA  & 0.6, 0.9999   \\
                            &        & Context noise $\ln\sigma$ & $\mathcal{N}(-1.2, 1.2^2)$ & $\alpha$ & $10^{-6}$ \\
                            &        & Clean-context prob.       & 0.25     & $s_0$                   & 10            \\
                            &        & Hist. noise (inference) & 0.3      & $s_1$                   & 1280          \\
                            &        &                           &          & $c$                     & 0.27648       \\
    \bottomrule
  \end{tabular}
  \label{tab:cm_training}
\end{table}

\begin{algorithm}
\caption{Channel-wise downscaling}
\begin{algorithmic}[1]
\Require Consistency model $\mathbf{f}_\theta$, interpolated and bias-corrected GCM field $\mathbf{y}$, ordered channel-wise noise scales $\sigma_C > \sigma_{C-1} > \cdots > \sigma_1 > \sigma_{\text{min}}$, $C=N_{\text{var}}$ is the total number of variable channels.
\State $\mathbf{\epsilon} \sim \mathcal{N}(\mathbf{0}, \mathbf{I})$
\State $\mathbf{y}_{C} \gets \mathbf{y} + \mathbf{\epsilon}\sigma_C $ \Comment{Forward noising}
\State $\mathbf{x} \gets \mathbf{f}_\theta(\mathbf{y}_{C}, \sigma_C)$ \Comment{Generative denoising}
\For{$i = C-1$ \textbf{to} $1$}
    \State $\mathbf{x}^c \gets \mathbf{y}^c$ for all $c \in [1,i]$ \Comment{Reset remaining channels}
    \State $\mathbf{\epsilon} \sim \mathcal{N}(\mathbf{0},\mathbf{I})$
    \State $\mathbf{x}_{i} \gets \mathbf{x} + \sqrt{\sigma_i^2 - \sigma_{\text{min}}^2} \, \epsilon$ \Comment{Forward noising}
    \State $\mathbf{x} \gets \mathbf{f}_\theta(\mathbf{x}_{i}, \sigma_i)$ \Comment{Generative denoising}
\EndFor
\State \Return $\mathbf{x}$
\end{algorithmic}
\label{alg:downscaling}
\end{algorithm}

\subsubsection{Multivariate Downscaling}
\label{sec:method_downscaling}

In order to determine the noise scale, we compute the spatial power spectral density (PSD) for each of the 8 variables in the standardized historical GCM fields. The noise scales $\sigma_i,$ $i=1,...,8$, are determined from the PSD at the wavenumber that corresponds to twice the GCM spatial resolution (i.e. Nyquist frequency) following \cite{hessFastScaleadaptiveUncertaintyaware2025}, and given in Table \ref{tab:noise_scale}.
We use multi-step CM sampling \cite{songImprovedTechniquesTraining2024} to denoise the different variables with individual noise scales (Alg.~\ref{alg:downscaling}), where we have omitted the inputs from current and previous time steps and static inputs. 
The sampler first denoises all variables using the largest noise (spatial) scale $\sigma_C$. 
Then all variables with a smaller noise (spatial) scale $\sigma_i < \sigma_C$ are restored from the GCM field, before the next denoising is performed again with the next smaller noise scale. We find small improvements by adding three additional denoising steps at the end for the smallest scales using the standard CM sampler \cite{songImprovedTechniquesTraining2024}.

\begin{table}[t]
\centering
\caption{Noise scales $\sigma_i$ used for downscaling per general circulation
  model and variable, in standardized units of the (transformed) model variables.
  $\lambda^\ast$ is the wavelength at which the injected noise matches the GCM
  spectral power.}
\begin{tabular}{l c c c c c c c c c}
\toprule
GCM & {$\lambda^\ast$ (km)} & pr & tas & tasrange & tasskew & sfcWind & rlds & rsds & hurs \\
\midrule
CESM1-CAM5      & 210 & 1.94 & 1.68 & 4.44 & 6.43 & 3.27 & 1.31 & 0.77 & 1.96 \\
HadGEM3-GC31-MM & 120 & 0.97 & 0.75 & 2.22 & 3.45 & 1.37 & 0.42 & 0.44 & 0.90 \\
MPI-ESM1-2-HR   & 200 & 2.62 & 1.36 & 3.81 & 6.47 & 2.49 & 1.25 & 0.99 & 1.67 \\
\bottomrule
\end{tabular}
\label{tab:noise_scale}
\end{table}

\section{Data Records}
In this section, we explain what the dataset contains, give an overview of the data files, their formats, and folder structure.

An overview of the dataset is given in Table \ref{tab:dataset_overview}. 
The folder structure is \texttt{gcm/scenario/file.nc} and each variable is stored as a single NetCDF file for the full scenario.
We follow standard CMIP file-naming conventions with \\ \texttt{var\_day\_esm\_scenario\_realization\_downscaling\_ML\_yyyymmdd--yyyymmdd.nc}, where the daily temporal resolution is denoted by \texttt{day}, while the start and end date are given at the end of the file name. 
Each NetCDF file follows the Climate and Forecast (CF) metadata conventions and contains the \texttt{time}, \texttt{lat} and \texttt{lon} coordinate variables, together with variable-level attributes (\texttt{standard\_name}, \texttt{long\_name} and \texttt{units}) and global attributes describing the source GCM, scenario, downscaling method and processing version.
The dataset comprises the downscaled GCM simulations without the raw GCM output. 
We apply level-5 DEFLATE compression, which results in file sizes between 74 GB and 109 GB. The total compressed dataset size is 1.6 TB.

\section{Technical Validation}
We evaluate ClimTip-GML using a comprehensive set of metrics, focusing on bias reduction, downscaling, multivariate correlations and the preservation of the climate change response in the GCM simulations.
We report results based on the last 10 years of each historical GCM simulation (1996 -- 2005 for \cesm and 2005 -- 2014 for \mpi and \hadgem) and compare them against the respective periods in ERA5. The time range choice reflects a balance between using data sets that are as independent as possible from the CM training data while accounting for the different simulation periods of the GCMs. 

\begin{figure}
    \centering
    \includegraphics[width=1.0\linewidth]{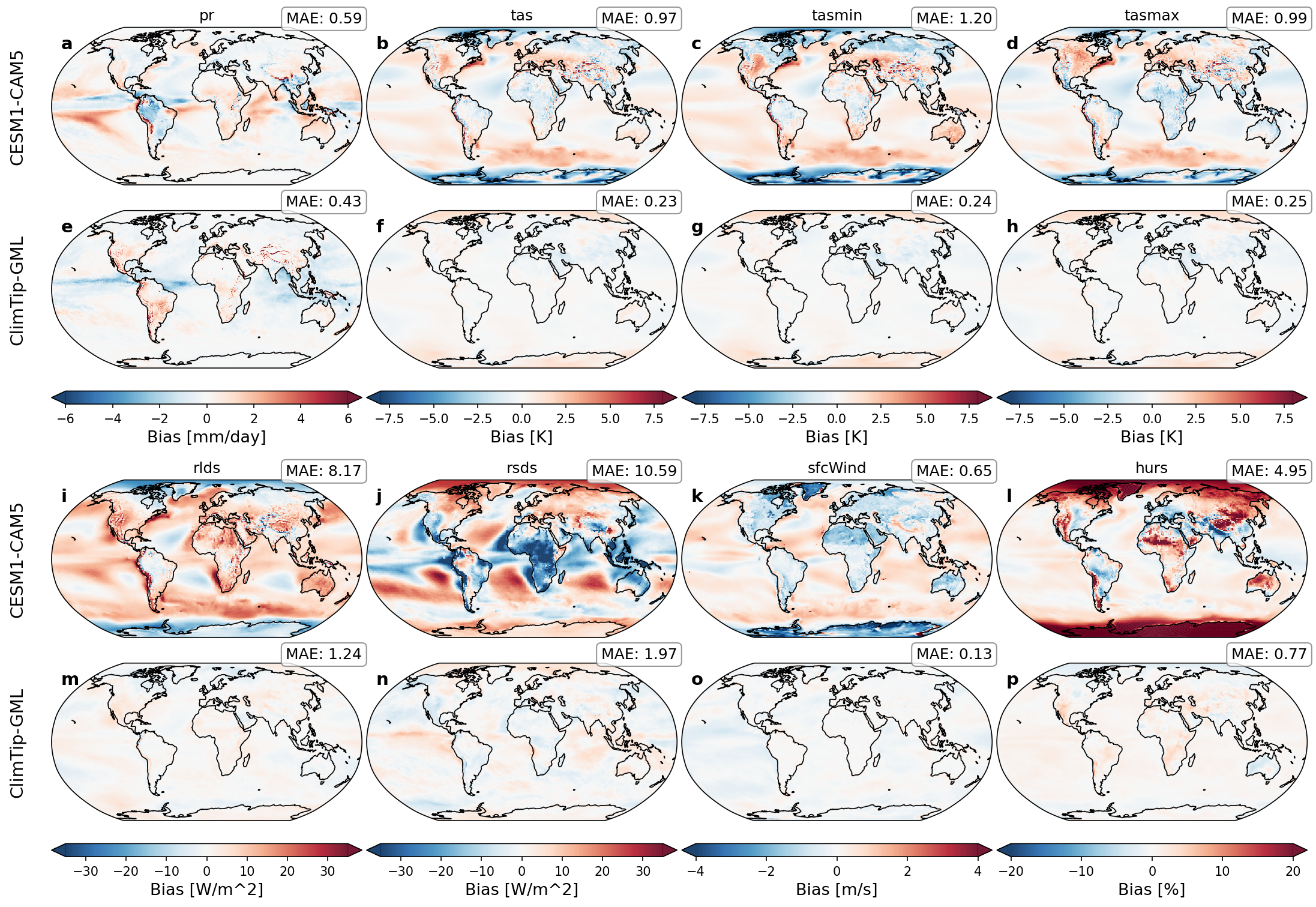}
    \caption{\textbf{Evaluation of bias adjustment for \cesm.} The difference to ERA5 in the 10-year-mean for 1996-2005, computed at each grid cell, is shown for all eight variables in the raw GCM historical output (\textbf{a} -- \textbf{d} and \textbf{i} -- \textbf{l}) and ClimTip-GML (\textbf{e} -- \textbf{h} and \textbf{m} -- \textbf{p}) of \cesm. The global mean absolute error (MAE) is reported in the top right corner of each panel. An effective reduction in long-term biases with considerably reduced spatial heterogeneity in the remaining error, can be seen across all variables.}
    \label{fig:bias_cesm}
\end{figure}

\begin{table}
 \caption{ClimTip-GML dataset overview. $^*$Note the stableT-2K-ARF scenario is only available for \cesm and \mpi.}
  \centering
  \begin{tabular*}{\textwidth}{@{\extracolsep{\fill}}l|lll}
    \toprule
    \textbf{Parameters}      & \multicolumn{3}{l}{\textbf{Specification}}     \\
    \midrule
    Variables               & \textit{Name}& \textit{Description}  & \textit{Units}\\
                            & \texttt{pr}     &  Total precipitation   &  mm/day                \\
                            & \texttt{tas}    &  Near-surface mean air temperature   &  K\\
                            & \texttt{tasmin} &  Near-surface min air temperature     & K     \\
                            & \texttt{tasmax} &  Near-surface max air temperature      & K     \\
                            & \texttt{rsds} &  Surface downwelling shortwave radiation   &  W/m$^2$      \\
                            & \texttt{rlds} &  Surface downwelling longwave radiation      &   W/m$^2$    \\
                            & \texttt{sfcWind} &  Near-surface wind speed  &  m/s\\
                            & \texttt{hurs} &  Near-surface relative humidity   &  \%        \\
    GCMs                    & \multicolumn{3}{l}{ \cesm, \hadgem, \mpi} \\
    Temporal resolution     & \multicolumn{3}{l}{daily}  \\
    Temporal coverage       & \multicolumn{3}{l}{100 years}       \\
    Spatial  resolution     & \multicolumn{3}{l}{0.25$^\circ \times$ 0.25$^\circ $}       \\
    Spatial  coverage       & \multicolumn{3}{l}{global}  \\
    Scenarios               & \multicolumn{3}{l}{\makecell[l]{piControl, historical, stableT-2K, stableT-2K-AMOC, and \\ stableT-2K-ARF$^*$}} \\
    File format & \multicolumn{3}{l}{NetCDF} \\
    File size & \multicolumn{3}{l}{74 GB -- 109 GB}   \\
    File number &  112 &\\
    \bottomrule
  \end{tabular*}
  \label{tab:dataset_overview}
\end{table}
\subsection{Bias Correction Validation}

\begin{figure}
    \centering
    \includegraphics[width=1.0\linewidth]{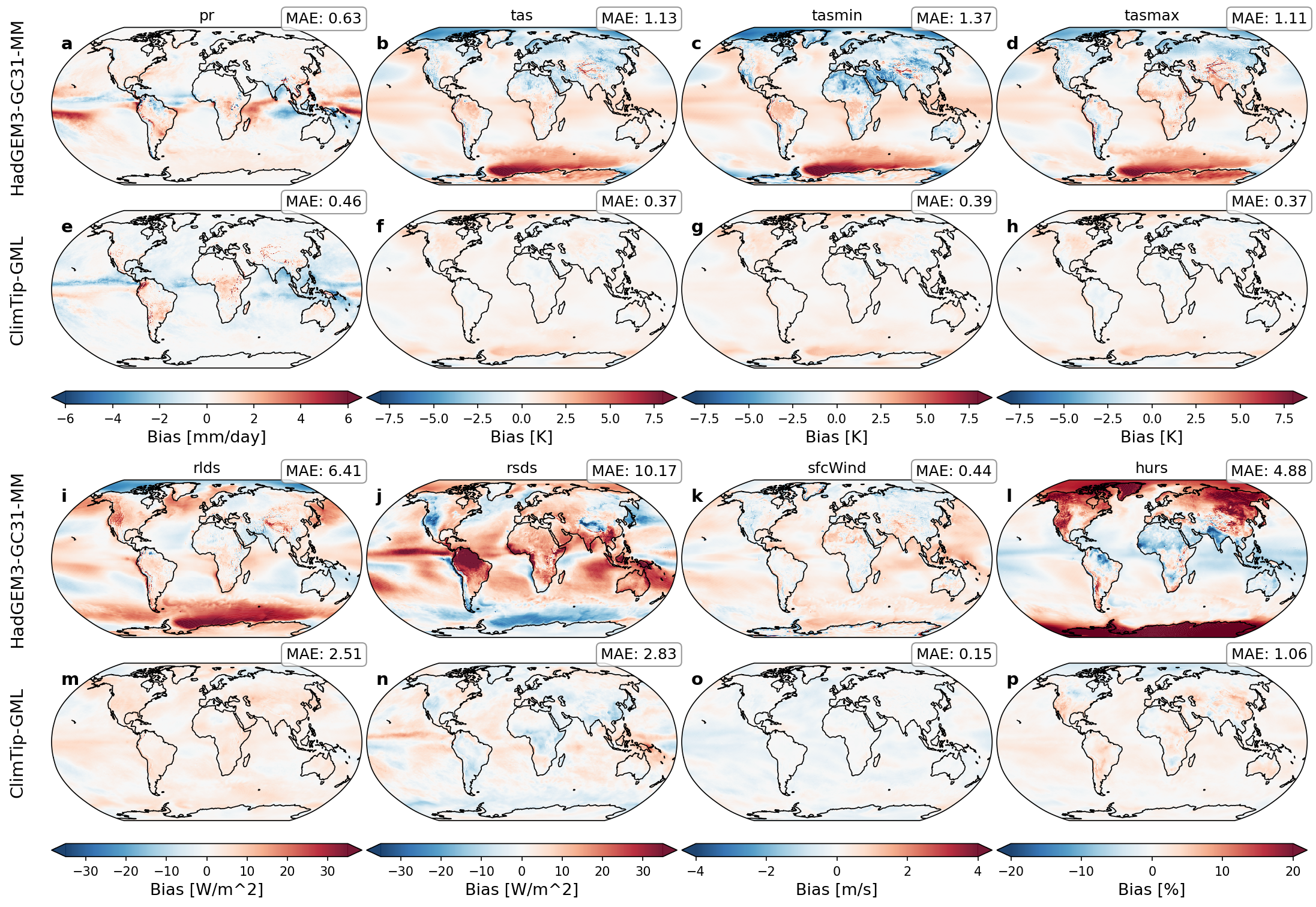}
    \caption{Same as Fig.~\ref{fig:bias_cesm}, but for \hadgem.}
    \label{fig:bias_hadgem}
\end{figure}

\begin{figure}
    \centering
    \includegraphics[width=1.0\linewidth]{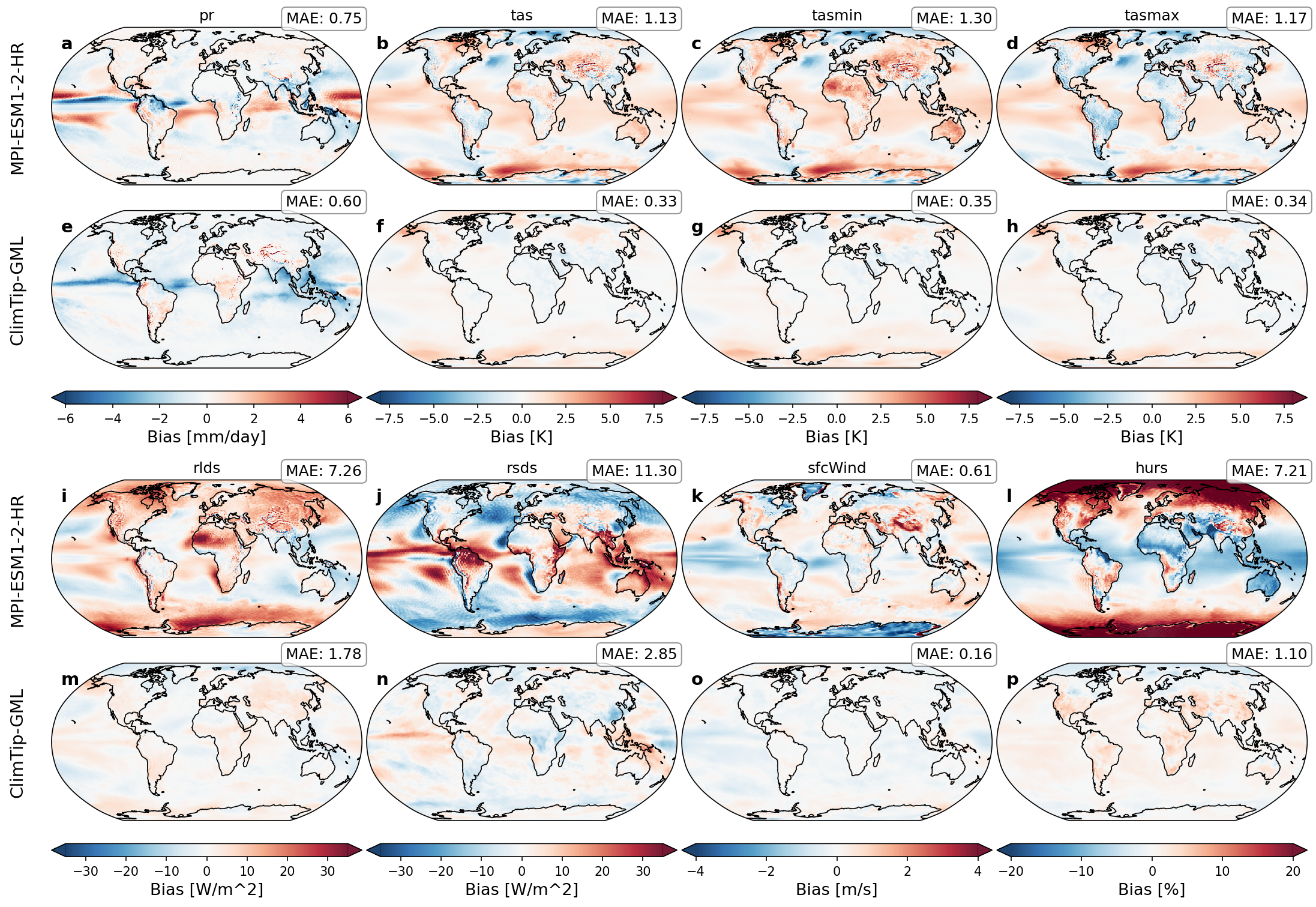}
     \caption{Same as Fig.~\ref{fig:bias_cesm}, but for \mpi.}
    \label{fig:bias_mpi}
\end{figure}

We compare biases in the 10-year mean of the raw GCMs and our post-processing in Fig.~\ref{fig:bias_cesm}, \ref{fig:bias_hadgem}, and \ref{fig:bias_mpi} for all eight variables, where the bias is computed at each location as
\begin{equation}
    \text{Bias} := \frac{1}{N} \sum_{n=1}^N (y^n -  x^n),
    \label{eq:bias}
\end{equation}
where $y^n$ and $x^n$ are single variables at a given grid cell from the (post-processed) GCM output and ERA5 respectively and the time index is denoted with $n$. The global weighted mean absolute error (MAE) is reported in the top right corner of each panel.\\ 
Overall the validation shows a reduction of long-term biases in ClimTip-GML compared to the raw GCMs for all three GCMs (Fig.~\ref{fig:bias_cesm} -- \ref{fig:bias_mpi}).
The MAE reductions are substantial, averaging 68\% across the full set of variables and GCMs.
The MAE bias reduction per model averaged over variables spans 64\% (for \hadgem) to 74\% (for \cesm).
Relative humidity (\hurs) shows the largest improvement, with an MAE reduction of 83\% on average.
The temperature variables (\tas, \tasmin, \tasmax) and radiation variables (\rlds, \rsds) and near-surface wind speed (\sfcwind) form an intermediate group, with average MAE reductions of 71 -- 76\% (e.g. \tasmin from 1.30 to 0.35 K, \rsds from 11.30 to 2.85 W/m$²$  and sfcWind from 0.61 to 0.16 m/s for \mpi).
Precipitation (\pr) exhibits the smallest relative improvement (25\% on average, from 0.59 -- 0.75 to 0.43 -- 0.60 mm/day), being the most challenging variable with the intermittent, heavy-tailed nature of the spatio-temporal fields.
The remaining bias is spatially homogeneous for all variables except precipitation, which retains a coherent dry bias along the tropical oceanic rain bands and a wet bias over tropical land.
Overall, these results validate that ClimTip-GML has effectively reduced biases compared to the raw GCM output across all variables.

\subsection{Spatial Downscaling Validation}

\begin{figure}
    \centering
    \includegraphics[width=0.935\linewidth]{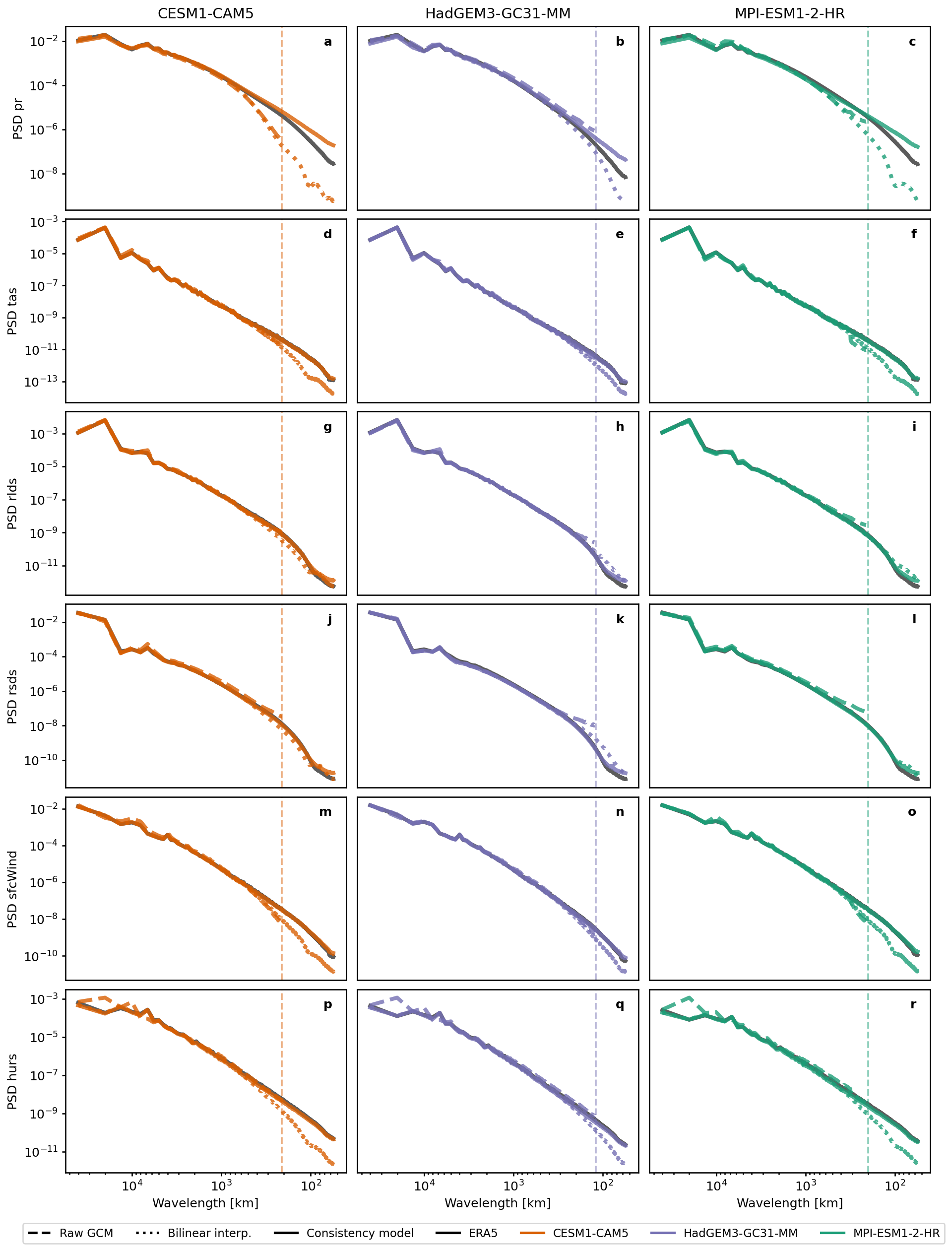}
    \caption{\textbf{Downscaling validation across GCMs and variables.}
    Spatial power spectral density (PSD) is shown for six variables (rows) and three GCMs (columns). Each panel compares the raw GCM (dashed colored), bilinear interpolation (dotted colored), and consistency model (solid colored) against the ERA5 reference (solid black). The vertical dashed line indicates the wavelength corresponding to twice the native GCM resolution.
    ClimTip-GML matches the ERA5 spectrum down to the smallest resolved scales ($\approx$ 28 km), with a small-scale power overestimation for precipitation.}
    \label{fig:psd}
\end{figure}

A skillful downscaling method adds realistic variability at spatial scales smaller than the native GCM grid while preserving patterns on the (bias-corrected) large scales.
We validate this by comparing the spatial power spectral density (PSD) averaged over all fields for each simulation and variable.
We choose six representative variables (\pr, \tas, \rlds, \rsds, \sfcwind, and \hurs) from all three GCMs, and report results from different stages of our processing pipeline: the raw GCM, bilinear interpolation and the generative consistency model-downscaling (Fig.~\ref{fig:psd}). 
The vertical dashed lines in Fig.~\ref{fig:psd} mark the wavelength corresponding to twice the native GCM resolution, i.e. the smallest scale that can be resolved by the raw GCM simulation.
The wavelength at this spatial scale also determines the noise scale in the downscaling method (see Alg.~\ref{alg:downscaling}). \\
At wavelengths larger than this scale, the raw GCM, bilinear interpolation, and consistency model-based downscaling all closely align with the ERA5 spectrum, confirming that neither the interpolation nor the generative downscaling distort the large-scale variability. 
At smaller scales, however, bilinear interpolation exhibits a steep drop in spectral power toward the grid scale ($\approx$ 28 km), reflecting its expected inability to generate small-scale subgrid variability, resulting in overly smooth fields.
The consistency model, by contrast, reconstructs the missing spectral power and closely follows the ERA5 PSD down to the smallest resolved wavelengths, recovering realistic small-scale variability for every GCM. 
For precipitation, the downscaled fields contain more power than ERA5 at wavelengths below the native GCM resolution.
Minor deviations remain at the smallest scales for the radiation and wind speed variables, but these are small relative to the large errors of the bilinear interpolation step.
The downscaling skill is consistent across the three GCMs despite their differing native average resolutions (\cesm $\approx$ 105 km, \hadgem $\approx$ 60 km, \mpi $\approx$ 100 km), indicating that the scale-adaptive noise conditioning (Section~\ref{sec:method_downscaling}) adds variability appropriate to each source.

\begin{figure}
    \centering
    \includegraphics[width=1.0\linewidth]{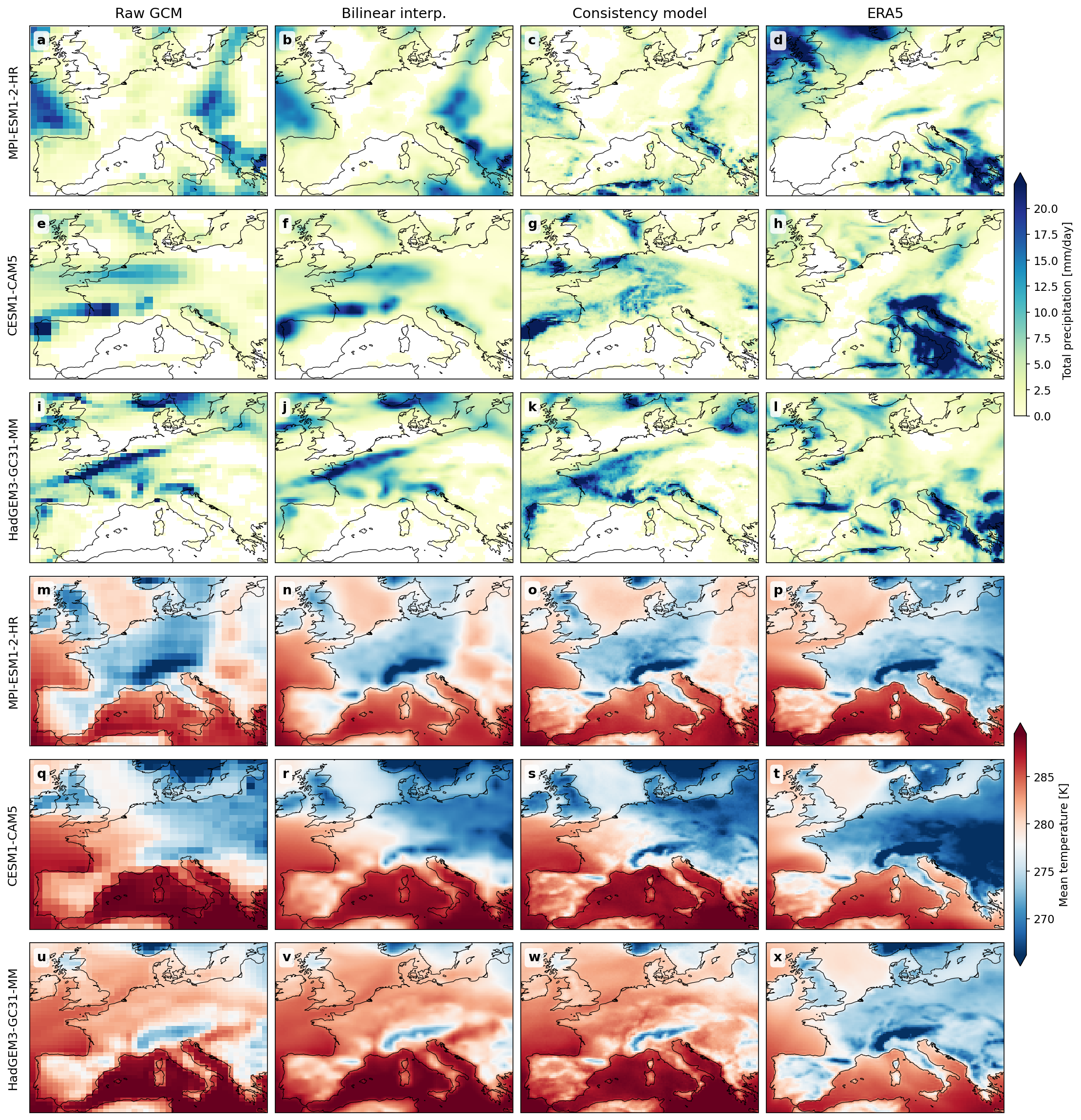}
    \caption{\textbf{Single precipitation and temperature fields over Europe.}
     Single precipitation and temperature fields are shown for different methods (columns) and three GCMs (rows). 
     The left column shows the raw GCM output, followed by bilinear interpolation, and consistency model downscaling from the a randomly selected day (27 December in the last validation year) as well as an unpaired ERA5 field for qualitative comparison.
     }
    \label{fig:regional_fields}
\end{figure}

Qualitative evaluation of randomly chosen single fields over a region covering western Europe is shown in Fig.~\ref{fig:regional_fields} for precipitation and mean temperature. Consistent with the PSD analysis in Fig.~\ref{fig:psd}, a lack of spatial variability is apparent in the bilinearly downscaled GCM fields, in particular over mountainous regions. The consistency model applied on top of the interpolation recovers small-scale variability that can be seen in the ERA5 reference.

\subsection{Multivariate Validation}

\begin{figure}
    \centering
    \includegraphics[width=0.9\linewidth]{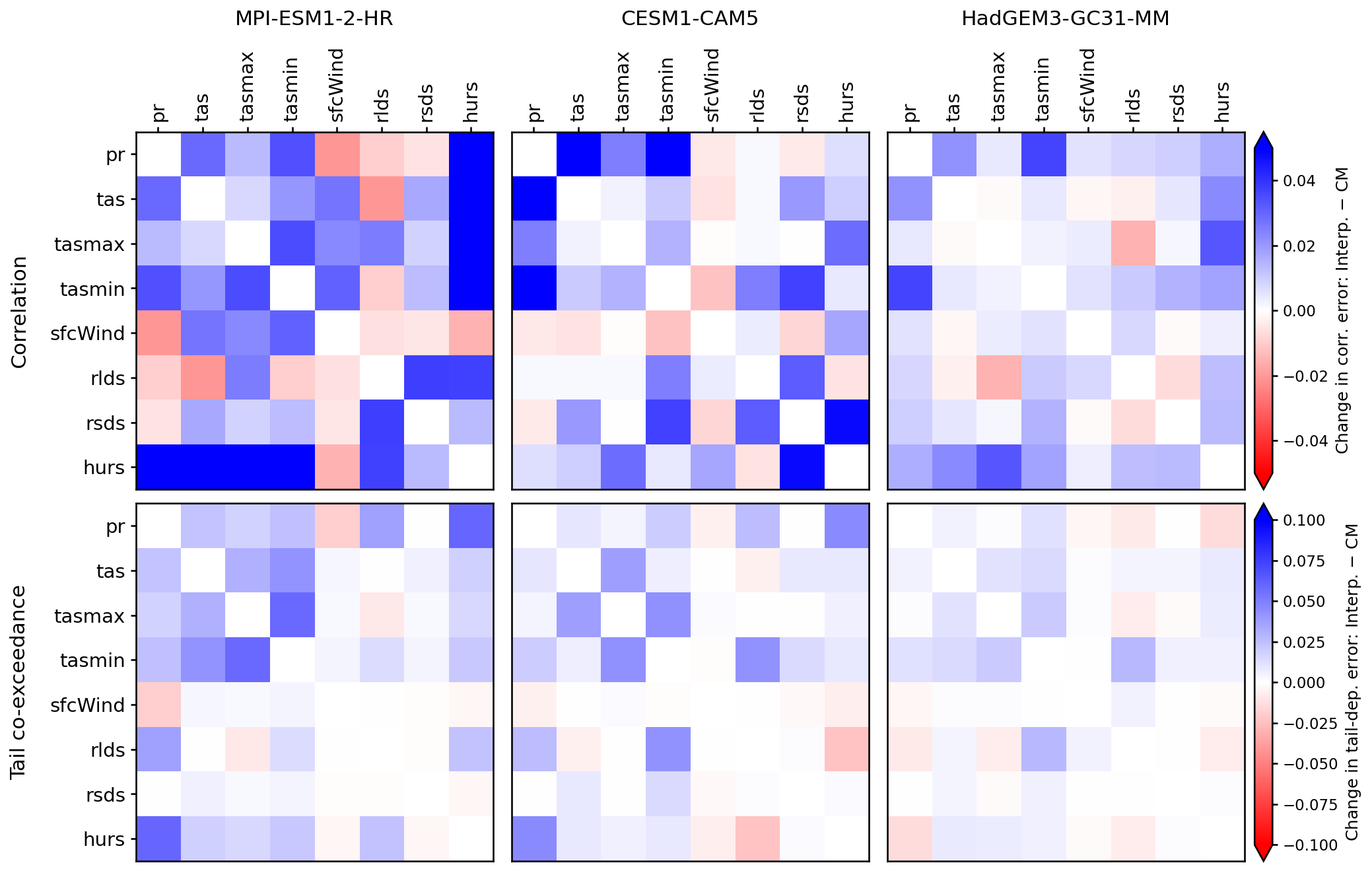}
    \caption{\textbf{Multivariate validation across GCMs.} Change in the error of pairwise multivariate dependencies for the three GCMs (columns: \mpi, \cesm, \hadgem). Rows show the Spearman rank correlation (top) and the tail co-exceedance probability (bottom) for all $8\times8$ variable pairs. Colors indicate the difference between the dependency error of the bilinear interpolation and of the consistency model downscaling, both applied to the QDM-corrected GCM relative to ERA5; positive (blue) values indicate an improvement by the consistency model.}
    \label{fig:multivariate_eval}
\end{figure}

We evaluate the multivariate dependency structure using two pairwise metrics computed for all $8\times8$ variable combinations: the Spearman rank correlation coefficient, characterizing the overall co-variability, and the tail co-exceedance probability $\chi_{i,j}(q)$, characterizing the joint occurrence of extremes above the $q=0.95$ quantile. 
The latter is defined following \cite{colesDependenceMeasuresExtreme1999a} as 
\begin{equation}
\chi_{ij}(q) \;=\; \Pr\!\left(F_j(X_j) > q \;\middle|\; F_i(X_i) > q\right) \;=
\; \frac{\Pr\!\left(F_i(X_i) > q,\; F_j(X_j) > q\right)}{1-q},
\label{eq:coexeedence}
\end{equation}
where $F_i$ and $F_j$ are cumulative distribution functions of two variables $X_i$, $X_j$.
We approximate Eq.~\ref{eq:coexeedence} using empirical quantiles $\hat q_i \;=\; \hat F_i^{-1}(q)$ with
\begin{equation}
\hat{\chi}_{ij}(q)
\;=\;
\frac{\frac{1}{N}\sum_{t=1}^{N}
      \mathbf{1}\!\left\{X_{i,t} > \hat q_i\right\}\,
      \mathbf{1}\!\left\{X_{j,t} > \hat q_j\right\}}
     {1-q}.
\end{equation}
For each variable pair we compute the absolute error of the metric relative to ERA5 and report the difference between bilinear interpolation and the consistency model downscaling (Fig.~\ref{fig:multivariate_eval}), such that positive values indicate a smaller error, i.e. an improved dependency, using the generative downscaling.\\
Across all three GCMs the difference is predominantly positive for the correlations, and for the tail co-exceedance in \mpi and \cesm, showing that the generative downscaling reproduces the inter-variable dependencies of ERA5 better than bilinear interpolation.
For \hadgem, the changes in tail co-exceedance are smaller and more mixed.
Relative to interpolation, the downscaling reduces the mean absolute inter-variable correlation error from 0.065 to 0.05 (23\%) and the tail co-exceedance error from 0.024 to 0.016 (35\%), averaged over all variable pairs and the three GCMs. These improvements are consistent across models, ranging from 15\% to 26\% for correlations and from 16\% to 46\% for tail dependence.
The correlation improvements are largest for pairs precipitation with temperature and humidity (\pr\ -- \tas, \pr\ -- \hurs) and for the humidity–temperature pairs (\hurs\ -- \tas, \hurs\ -- \tasmin, \hurs -- \tasmax), related to the physical coupling between moisture, temperature, and the surface energy balance. 
The tail co-exceedance errors are similarly reduced, most notably within the temperature variables (\tas, \tasmin, \tasmax) and for precipitation with humidity and temperature, and for \tasmin\ -- \rlds, indicating that the joint behavior of extremes is better captured.
Some of the pairs show a slight degradation (red cells), for example wind-related pairs and individual temperature -- radiation combinations, but these are limited in number relative to the overall improvements. 
The results demonstrate that the generative downscaling on top of bilinear interpolation adds physically consistent multivariate structure at the downscaled resolution across all three GCMs.

\subsection{Climate Response Validation}

\begin{figure}
    \centering
    \includegraphics[width=1.0\linewidth]{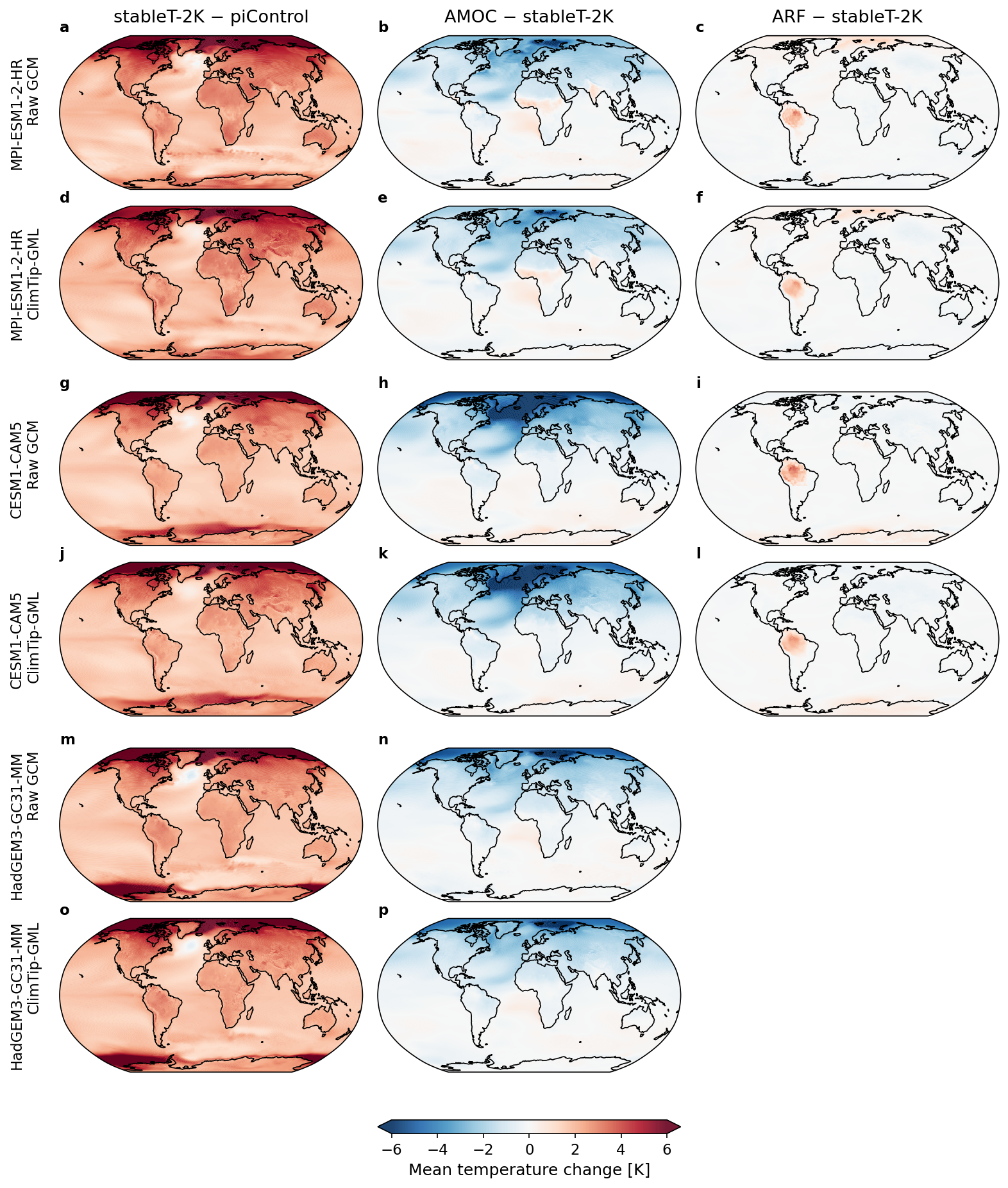}
    \caption{\textbf{Temperature climate change response validation.} Long-term change for mean daily temperature (TAS) [K] across the full 100-year scenarios (columns: stableT-2K $-$ piControl, AMOC $-$ stableT-2K, ARF $-$ stableT-2K) for the raw GCM and ClimTip-GML of each model (row pairs). The ARF tipping change is shown only for the models in which the experiment is available (\mpi and \cesm). The downscaled fields preserve the large-scale forced and tipping-induced responses of the raw GCM.}
    \label{fig:tip_tas}
\end{figure}

\begin{figure}
    \centering
    \includegraphics[width=1.0\linewidth]{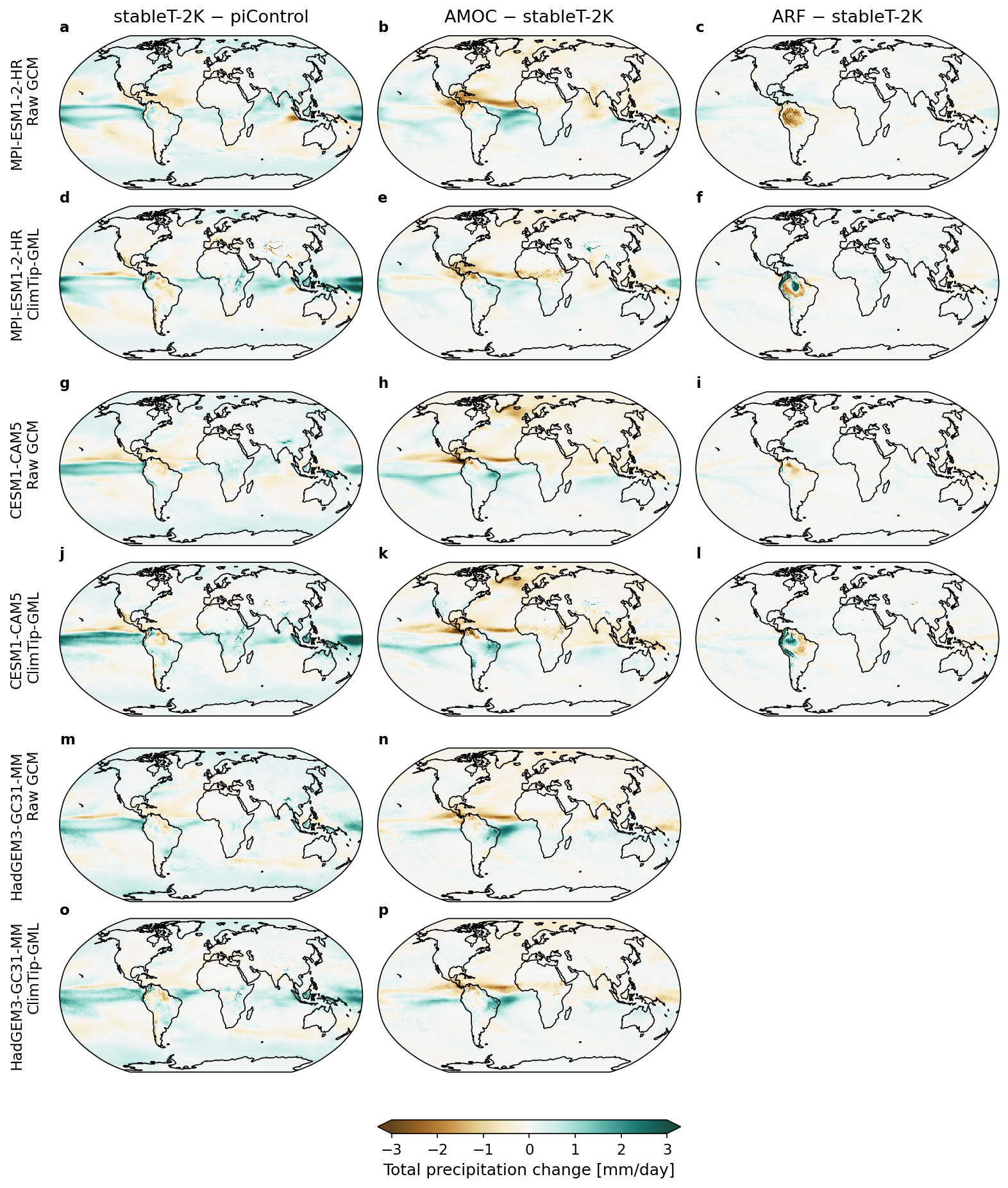}
    \caption{Same as Fig.~\ref{fig:tip_tas}, but for precipitation.}
    \label{fig:tip_pr}
\end{figure}

\begin{figure}
    \centering
    \includegraphics[width=1.0\linewidth]{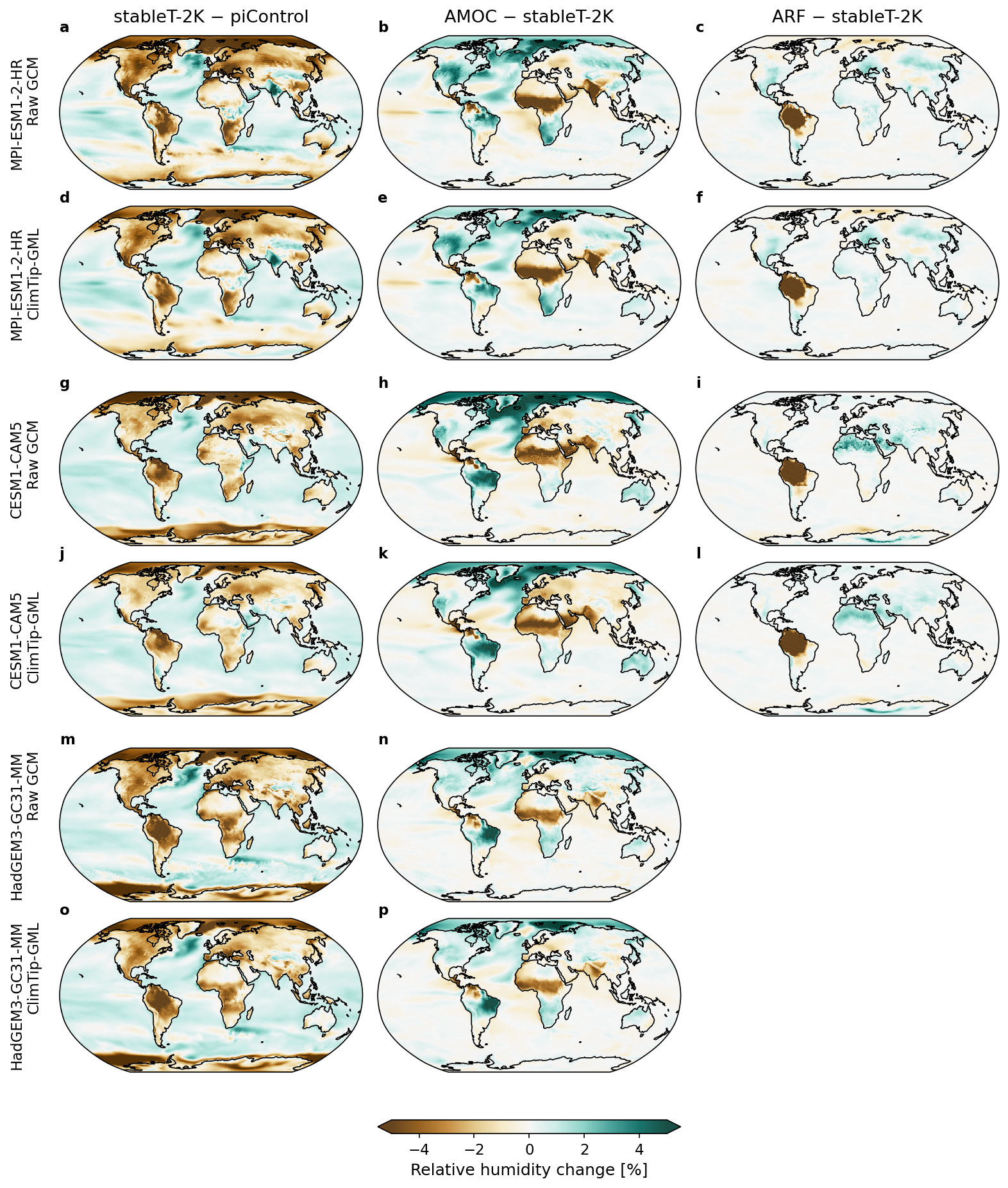}
    \caption{Same as Fig.~\ref{fig:tip_tas}, but for relative humidity.}
    \label{fig:tip_hurs}
\end{figure}

A bias correction and downscaling method for climate scenario analysis and impact assessment needs to preserve the large-scale climate response due to external forcing or tipping events.
We validate this for three representative variables, namely mean temperature (\tas, Fig.~\ref{fig:tip_tas}), total precipitation (\pr, Fig.~\ref{fig:tip_pr}), and relative humidity (\hurs, Fig.~\ref{fig:tip_hurs}), by comparing for each GCM the raw and ClimTip-GML fields across three cases: the stable 2K-warming signal relative to preindustrial control (stableT-2K $-$ piControl), the response to an AMOC collapse (AMOC $-$ stableT-2K) relative to the stable 2K warming, and similarly the response to an Amazon deforestation (ARF $-$ stableT-2K). 
ClimTip-GML reproduces the spatial pattern and magnitude of the corresponding raw GCM response. 
The 2K-warming signal shows the expected structure of land-amplified and polar-amplified warming, a moistening–drying redistribution of precipitation, and predominant continental relative-humidity reductions, all of which are retained after downscaling. 
The largest deviations from the raw GCM simulation are for precipitation in the tropics. 
Here, ClimTip-GML preserves the sign and pattern of the response but amplifies the precipitation increase along the equatorial Pacific and Maritime Continent and adds grid-scale structure over steep orography, e.g. the Himalaya.
The AMOC-collapse fingerprint is well-preserved, with the characteristic North Atlantic and Northern Hemisphere cooling (strongest in \cesm), the associated southward shift of tropical precipitation, and the accompanying relative humidity anomalies appearing consistently in both the raw and downscaled fields. 
The localized Amazon tipping signal is characterized by a regional warming together with a pronounced precipitation and relative-humidity decrease over tropical South America. 
The warming and the relative-humidity decrease are faithfully reproduced by the downscaling. The precipitation response, however, is not well preserved within the deforested region. While the raw GCMs show a precipitation decrease across the Amazon basin, ClimTip-GML shows a precipitation increase in the central and western basin and along the Andes, surrounded by a decrease. 
It is overall faithfully reproduced by the downscaling, with the largest difference again in the precipitation response over tropical South America, where an amplified decrease in precipitation over the Amazon basin as well as an increase over the Andes and the central South American continent is visible.

\section{Data Availability}
The ClimTip-GML dataset will be made publicly available for download. 
The ERA5 reanalysis is available at \url{https://cds.climate.copernicus.eu/datasets/reanalysis-era5-single-levels?tab=overview}. The monthly mean output of the GCMs is available at \url{https://zenodo.org/records/20324954}.

\section{Code Availability}
The Python code for entire project, including pre-processing, neural network training, post-processing and plotting will be available on GitHub at \url{https://github.com/p-hss/generative-downscaling-climtip.git}. The repository documentation includes comprehensive instructions on how to run the code, download the neural network weights and install required python packages used to ensure full reproducibility.

\bibliographystyle{naturemag}
\bibliography{references}  

\section*{Author Contributions}
PH and NB conceived the research and designed the study with
input from SB.
LC, LJ, and CP performed the GCM simulations.
PH performed the model training, downscaling and numerical analysis.
PH, SB, LC, LJ, CP, and NB interpreted and discussed the results.
PH wrote the manuscript with input from SB, LC, LJ, CP, and NB.

\section*{Competing Interests}
The authors declare no competing interests.

\section*{Acknowledgements}
This is ClimTip contribution \#223; the ClimTip project has received funding from the European Union's Horizon Europe research and innovation programme under Grant 101137601: Funded by the European Union. Views and opinions expressed are however those of the author(s) only and do not necessarily reflect those of the European Union or the European Climate, Infrastructure and Environment Executive Agency (CINEA). Neither the European Union nor the granting authority can be held responsible for them. 
PH is funded by the Deutsche Forschungsgemeinschaft (DFG, German Research Foundation) project number 556377191.
CESM simulations and analyses were performed on the Dutch National Supercomputing facilities, sponsored by NWO Exact and Natural Sciences under project 2024.017 (Anna von der Heydt) and 2024.013 (Henk Dijkstra).

\section*{Funding}
All funding is stated in the Acknowledgments section.

\section*{Ethics Statement}
This study does not involve human participants, personal data, or animal experiments; no ethical approval was therefore required.

\end{document}